# The Key Factors Controlling the Seasonality of Planetary Climate

Ilai Guendelman[1] and Yohai Kaspi[1]

[1]Earth and Planetary Department, Weizmann Institute of Science, Rehovot, Israel

**Abstract** Several different factors influence the seasonal cycle of a planet. This study uses a general circulation model and an energy balance model (EBM) to assess the parameters that govern the seasonal cycle. We define two metrics to describe the seasonal cycle, $\phi_s$, the latitudinal shift of the maximum temperature, and $\Delta T$, the maximum seasonal temperature variation amplitude. We show that alongside the expected dependence on the obliquity and orbital period, where seasonality generally strengthens with an increase in these parameters, the seasonality depends in a nontrivial way on the rotation rate. While the seasonal amplitude decreases as the rotation rate slows down, the latitudinal shift, $\phi_s$, shifts poleward. A similar result occurs in a diffusive EBM with increasing diffusivity. These results suggest that the influence of the rotation rate on the seasonal cycle stems from the effect of the rotation rate on the atmospheric heat transport.

**Plain Language Summary** The seasonal cycle that a planet undergoes can be described by its amplitude, that is, the maximal temperature difference experienced across the year. In addition, it can be characterized by the latitudinal shift of the maximum temperature during that annual cycle. This study explores the role of obliquity, orbital period, and rotation rate on the seasonality. We find that, as expected, increasing the obliquity and orbital period results in a stronger seasonal cycle. Surprisingly, decreasing the rotation rate has a complex effect on the seasonal cycle. It reduces the seasonal cycle amplitude but increases the latitudinal shift of the maximum temperature. We find that this is a result of the effect of the rotation rate on the way heat is redistributed within the atmosphere. These results emphasize the importance of studying the climate's dependence on planetary parameters, as for example, an Earth-like planet with a slower rotation rate will experience a significantly different seasonal cycle.

## 1. Introduction

The seasonal cycle that a planet experiences is a function of a large number of parameters. A necessary condition for a planet to experience a seasonal cycle is to have either non-zero obliquity or non-zero eccentricity, such that the incoming solar radiation will have a seasonal cycle. The degree of seasonality, however, depends on other parameters (e.g., Guendelman & Kaspi, 2019, 2020). The solar system's terrestrial planetary bodies, Earth, Mars, and Titan, are great examples of planets with a similar obliquity, having a similar seasonal pattern of the incoming solar radiation at the top of their atmosphere. Still, each planet experiences a different temperature seasonal cycle at its surface (Figure 1). This is a result of the different characteristics of these planets, especially, their radiative timescale. The atmospheric radiative timescale can be estimated by (Mitchell & Lora, 2016)

$$\tau_R = \xi_{IR} \frac{C_p P}{4 g \sigma T_e^3}, \tag{1}$$

where $\xi_{IR}$ is the infrared optical depth, $C_p$ is the specific heat at constant pressure, $P$ is pressure, $g$ is surface gravity (with $P/g$ is the atmospheric mass per unit area), $\sigma$ is the Stefan-Boltzmann constant, and $T_e$ is the equilibrium temperature. To leading order the ratio between the radiative timescale and the orbital period should dictate the strength of the seasonality that a planet will experience. For example, a planet with an orbital period that is significantly longer than its radiative timescale, will experience a strong seasonal cycle, and vice versa (e.g., Guendelman & Kaspi, 2019).

Earth has an atmospheric radiative timescale of ∼30 days (Wells, 2011). Additionally Earth's surface is mostly covered by deep oceans resulting in high surface thermal inertia. The combination of high surface thermal inertia and long atmospheric radiative timescale contribute to the strong modulation of Earth's seasonal cycle. Mars, on the other hand has a dry rocky surface with low thermal inertia and a thin atmosphere (e.g., Read et al., 2015).

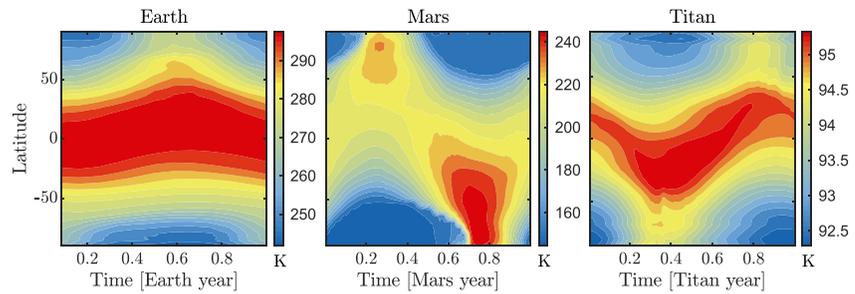

**Figure 1.** The seasonal cycle of the zonally averaged surface temperature on Earth (climatology from 1980 to 2019, reanalysis data from ERA5), Mars (climatology from reanalysis data from Greybush et al., 2019) and Titan (from a Titan atmospheric model used in Faulk et al. (2020)).

With an atmospheric radiative timescale of ∼1 day (Wells, 2011), it features the strongest seasonal cycle among the three. In addition, Mars's eccentric orbit results in an hemispheric asymmetry between the southern and northern summer solstices (Figure 1). Titan, conversely, has a thick, cold atmosphere, that is, a long radiative timescale (∼200 years, Mitchell & Lora, 2016), that also absorbs significant amounts of the shortwave radiation. All these factors modulate its seasonal cycle. However, it has a very low surface thermal inertia, which acts to increase the seasonal cycle. The competition between these two effects results in Titan having a moderate seasonal cycle. Another important parameter that differs among the three planets is the rotation rate, with Earth and Mars being fast rotating planets and Titan being a slowly rotating planet (with a rotation period of ∼16 Earth days). Previous studies have shown that the atmospheric heat transport (AHT) on slowly rotating planets is more efficient, resulting in flatter temperature gradients (Edson et al., 2011; Kaspi & Showman, 2015; Liu et al., 2017; Merlis & Schneider, 2010; Noda et al., 2017; Pierrehumbert & Hammond, 2019). This in turn has the potential to change the seasonal cycle that a planet experiences (Tan, 2022).

The obliquity of the giant planets, Saturn, Neptune, and Uranus, too, is significantly larger than zero. When considering their long atmospheric radiative timescale, most studies agree that these planets should experience weak-to-no seasonal cycle (e.g., Conrath et al., 1990). However, Li et al. (2018) have estimated a shorter radiative timescale for the ice giants, suggesting that both Uranus and Neptune should experience some seasonality. At least for Uranus, where the dynamics extends deep (Kaspi et al., 2013) this is in contrast to what current observations suggest (Orton et al., 2015; Roman et al., 2020). An opposite example of a mismatch between the observed and expected seasonality is Titan. The cold conditions and massive atmosphere of Titan result in an atmospheric radiative timescale that is much longer than its orbital period (Mitchell & Lora, 2016). However, the seasonal cycle on Titan is not negligible, and in some aspects stronger than that of Earth (Figure 1). Mitchell and Lora (2016) have suggested that the strength of the seasonal cycle is forced from Titan's low surface thermal inertia. Shortwave radiation warms the surface, which in turn radiates back to the atmospheric boundary layer that has a shorter radiative timescale compared to the entire atmosphere, allowing the seasonal cycle to become significant. Although this can explain the seasonality on Titan, in this study, we also explore the role of the rotation rate and AHT on the seasonal cycle.

This diversity in the seasonality of the solar system planets is tiny compared to the potential variety among planets outside the solar system. Even if remaining restricted to terrestrial planets within the habitable zone of their host star, the possible diversity is immense. For example, differences between ocean and rocky worlds, planets with different obliquities, eccentricities, or atmospheric compositions that would affect both the short-and long-wave optical depths, all these have an effect on the climate seasonality that a planet will experience (e.g., Armstrong et al., 2014; Colose et al., 2019; Guendelman et al., 2021, 2022; Linsenmeier et al., 2015; Lobo & Bordoni, 2020). In addition to all these, recent studies have also shown that the planetary rotation rate has an effect on the surface temperature latitudinal distribution and seasonal cycle (e.g., Edson et al., 2011; Guendelman & Kaspi, 2019; Kaspi & Showman, 2015; Liu et al., 2017).

This study explores the seasonality dependence on different planetary parameters in an idealized setting, connecting results from a simplified general circulation model (GCM) and a diffusive energy balance model (EBM). In the idealized GCM we vary the obliquity, orbital period, and rotation rate. The obliquity represents the role of the solar forcing, the orbital period represents the role of the atmospheric radiative response, and the rotation rate represents the role of the dynamics. Although, as mentioned, numerous other parameters can also influence the temperature, we consider these three as each represents a different part of the radiative and dynamical response of the climate system. Similarly, in the EBM we explore the roles of obliquity and of a non-dimensional radiative timescale and diffusion. As in the parameters studied in the GCM, these parameters have similar significance, with the non-dimensional radiative timescale representing the ratio between the orbital period and radiative timescale and the diffusion reflecting the role of the dynamics and AHT.

Note that in this study we explore the climate of planets in a circular orbit, that is, planets with zero eccentricity. Planets in an eccentric orbit will experience a more complex seasonal cycle, which can include hemispheric asymmetry, with different hemispheres experiencing different seasonal intensity (Guendelman & Kaspi, 2020; Ohno & Zhang, 2019) and is outside the scope of this study. This manuscript is arranged as follows, in Section 2, we describe the GCM and EBM used in this study. Section 3 describes the results, and we conclude and discuss the significance of this study in Section 4.

## 2. Methods

### 2.1. General Circulation Model

In this study, we use an idealized GCM with a seasonal cycle (Guendelman & Kaspi, 2019). The top of the atmosphere is forced by diurnal mean insulation, given by:

$$Q = \frac{S_0}{\pi} \left( h \sin \phi \sin \delta + \cos \phi \cos \delta \sin h \right), \tag{2}$$

where $S_0$ is the solar constant, $h$ is the hour angle of the sun at sunrise and sunset, $\phi$ is for latitude and $\delta$ is the declination angle (Hartmann, 2016). The model is an aquaplanet model with a 10 m depth mixed layer as the lower boundary. The model solves the primitive equations in a T42 horizontal resolution and 25 vertical layers. The radiation in the model is represented by a two-stream gray radiation scheme. The model utilizes a simplified parameterization for moist convection processes (Frierson et al., 2006).

We vary the parameters as follows:

- Obliquity ($\gamma$): 10, 20, 30, 40, 50, 60, 70, 80, 90.
- Orbital period ($\omega$): $\frac{1}{8}, \frac{1}{4}, \frac{1}{2}, 1, 2, 4$.
- Rotation rate ($\Omega$): $\frac{1}{16}, \frac{1}{8}, \frac{1}{4}, \frac{1}{2}, 1, 2$.

The values are normalized such that $\omega = 1$ is an orbital period of 360 Earth days, and $\Omega = 1$ is Earth's rotation rate. We vary two parameters at a time while keeping the third constant, with Earth-like values, except for the obliquity, which is 30°. This results in a total of ∼142 simulations. The Earth-like simulation (top right corner in Figure 2) has a stronger seasonal cycle compared to observations as it has a higher obliquity, a relatively low mixed layer depth, and due to the model's simplicity. Although planets in a short orbital period can become tidally locked, we do not consider this configuration in this study.

### 2.2. Energy Balance Model

A more simplified way to study the temperature response to the different parameters is a diffusive EBM, given by:

$$C \frac{\partial T}{\partial t} = Q(1 - a) - \sigma T^4 + \frac{\partial}{\partial \phi} \left( D \cos \phi \frac{\partial T}{\partial \phi} \right), \tag{3}$$

where $C$ is the effective heat capacity, with the value of $C$ for a 10 m mixed layer depth being ∼$4 \times 10^7$ Jm$^{-2}$ K$^{-1}$, $T$ is the temperature, $Q$ is the diurnal mean insolation, $a$ is the albedo, $\sigma$ is the Stefan-Boltzmann constant, $D$ is the diffusion coefficient, taken to be constant (an appropriate value for Earth is ∼0.6 Wm$^{-2}$ K$^{-1}$, North & Kim, 2017), and $\phi$ is latitude. The diurnal mean insolation, $Q$, is calculated using Equation 2. Although it is conventional to use a linear form for the outgoing longwave radiation (OLR, e.g., Budyko, 1969; North & Coakley, 1979; Rose et al., 2017), and is even found to be more accurate for Earth (Koll & Cronin, 2018), it involves some assumptions as it can vary among different planetary atmospheres depending on their composition, thus we use the blackbody radiation form for the OLR instead of the linear one. Additionally, Henry and Vallis (2021) have shown that the seasonality is significantly influenced from nonlinear effects of the blackbody radiation.

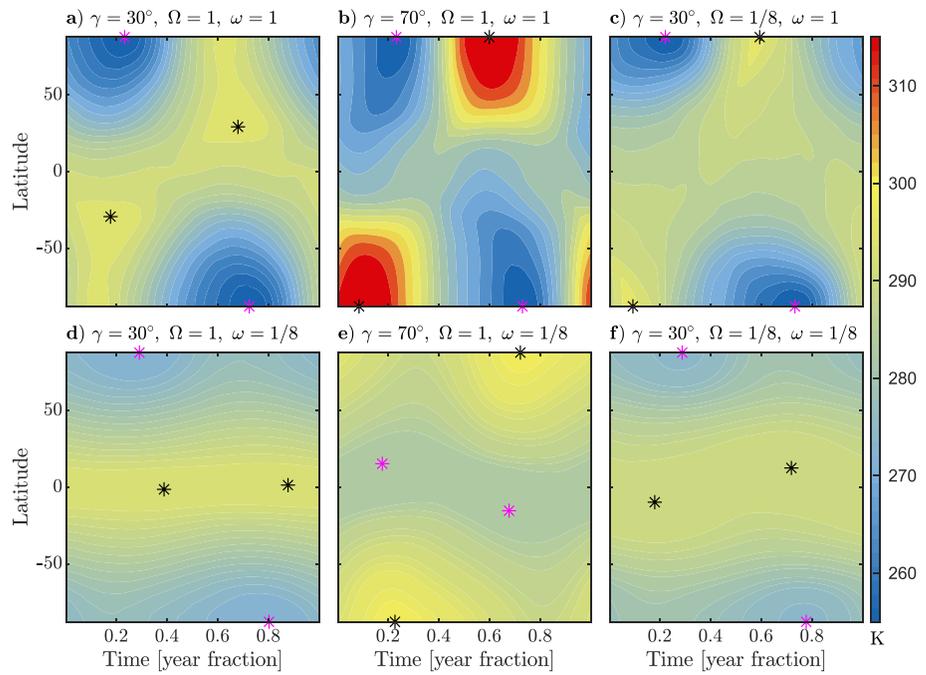

**Figure 2.** Zonally averaged surface temperature as a function of latitude and annual cycle for different simulations from the idealized general circulation model; the obliquity ($\gamma$), rotation rate ($\Omega$) and orbital period ($\omega$) values for each simulation are given in the title. Black (magenta) asterisk represent the maximum (minimum) temperature at each hemisphere. Southern (Northern) hemisphere summer solstice is in ∼0.97 (∼0.47, in year fraction units).

Even with these simplifications, the temperature seasonal cycle depends on a large number of variables, such as the heat capacity, solar constant, orbital period, obliquity, and diffusion. To reduce the number of variables, we use a non-dimensional form of Equation 3. Consider the following transformation:

$$T = \overline{T} T', \tag{4}$$

$$t = \omega t', \tag{5}$$

where

$$\overline{T} = \left( \frac{\overline{Q}}{\sigma} \right)^{1/4}, \tag{6}$$

with $\overline{Q} = S_0(1-a)/4$ denoting the annual, global mean insolation, and $\omega$ the orbital period. Substituting the transformation in Equation 3 gives:

$$\frac{C}{\sigma \overline{T}^3} \frac{1}{\omega} \frac{\partial T'}{\partial t'} = Q' - T'^4 + \frac{D}{\sigma \overline{T}^3} \frac{\partial}{\partial \phi} \left( \cos \phi \frac{\partial T'}{\partial \phi} \right). \tag{7}$$

Since $C/\sigma \overline{T}^3$ scales like the radiative timescale, $\tau_R$, we can define the following non-dimensional numbers.

$$\eta = \frac{C}{\sigma \overline{T}^3} \frac{1}{\omega}, \tag{8}$$

$$\mu = \frac{D}{\sigma \overline{T}^3}, \quad (9)$$

to write

$$\eta \frac{\partial T'}{\partial t'} = Q' - T'^4 + \mu \frac{\partial}{\partial \phi}\left(\cos\phi \frac{\partial T'}{\partial \phi}\right). \quad (10)$$

Here, the resulting seasonal cycle depends on three parameters, the obliquity ($\gamma$), the ratio between the radiative timescale and the orbital period ($\eta$), and the non-dimensional diffusion coefficient ($\mu$); a similar non-dimensional analysis was done for a linearized model in Rose et al. (2017). We solve Equation 10 using an explicit finite difference method (North & Kim, 2017), for different values of $\gamma$, $\eta$, and $\mu$. Note that using the annual, global mean insolation to non-dimensionalize Equation 3 is an approximation. This approximation neglects, for example, latitudinal differences in the radiative timescale, which can be important (Guendelman & Kaspi, 2020; Mitchell et al., 2014). However, for the purpose of this study, this approximation is sufficient.

### 2.3. Seasonality Indices

The climate seasonality can be considered to be the deviations of the climate from the annual mean climate, or in the case of an aquaplanet, deviations from hemispheric symmetry. There are several ways of quantifying the seasonality. Here we focus mainly on the latitudinal shift of the maximum temperature during the year, $\Delta(\phi_{max})$. Accounting for the two hemispheres, we use

$$\phi_s = \Delta(\phi_{max})/2, \quad (11)$$

to describe the seasonal cycle. Note that this definition becomes problematic for planets with obliquity higher than 54° and a relatively short orbital period (bottom row middle panel in Figure 2), as this combination will lead to an overestimation of the seasonality. In this case, the annual mean climate is reversed (e.g., Guendelman & Kaspi, 2019; Kang et al., 2019), with the maximum temperature located at the poles and the minimum temperature at the equator. As a result, for planets with a short orbital period and high obliquity, it is more appropriate to follow the latitudinal shift of the minimum temperature ($\Delta(\phi_{min})$) rather than that of the maximum one, as the behavior of the minimum temperature is similar to that of the maximum temperature in low obliquities, and will describe the seasonality more appropriately (bottom row in Figure 2). Formally, to account for this complexity, we redefine the seasonality index (Equation 11) to be

$$\phi_s = \min[\Delta(\phi_{max}), \Delta(\phi_{min})]/2. \quad (12)$$

This latituinal shift $\phi_s$ is the main focus of this study, and a crucial characteristic of the seasonal cycle. However, $\phi_s$ does not fully characterize the seasonal cycle, and we also examine the effect of the different parameters on the seasonal amplitude of temperature,

$$\Delta T = \max|T - \langle T \rangle|, \quad (13)$$

where $\langle T \rangle$ is the annual mean temperature such that $\Delta T$ describes the maximal deviation from the annual mean climate during the seasonal cycle. Together, $\Delta T$ and $\phi_s$, give a more complete picture of the seasonal cycle.

## 3. Results

### 3.1. General Circulation Model

The temperature seasonal cycle varies within the parameter space. Figure 2 shows six different cases that can qualitatively describe the effect of each parameter in the surface temperature seasonal cycle. The most straight-froward response is that decreasing the orbital period decreases the seasonality, that is, a decrease in both $\phi_s$ and $\Delta T$ (Figures 2a and 2d). For increasing the obliquity, we can distinguish two types of responses, depending on the orbital period. For a long enough orbital period, increasing the obliquity results in an increase in both $\phi_s$ and $\Delta T$ (Figures 2a and 2b). However, for short orbital periods, the response is more complex, where for high obliquities there is a reversal of the temperature gradients, that is, for high obliquities, maximum temperatures are at the poles instead of the equator during the entire seasonal cycle (Figure 2e). Note that the seasonality indices for high and low obliquity in a short orbital describe an inherently different climate, one with maximum temperatures at the equator and the other with maximum temperature at the pole, respectively (Figures 2d and 2e). Decreasing the rotation rate has a more complex response, where the seasonal amplitude, $\Delta T$, decreases while $\phi_s$ increases with the rotation rate (Figures 2a and 2c).

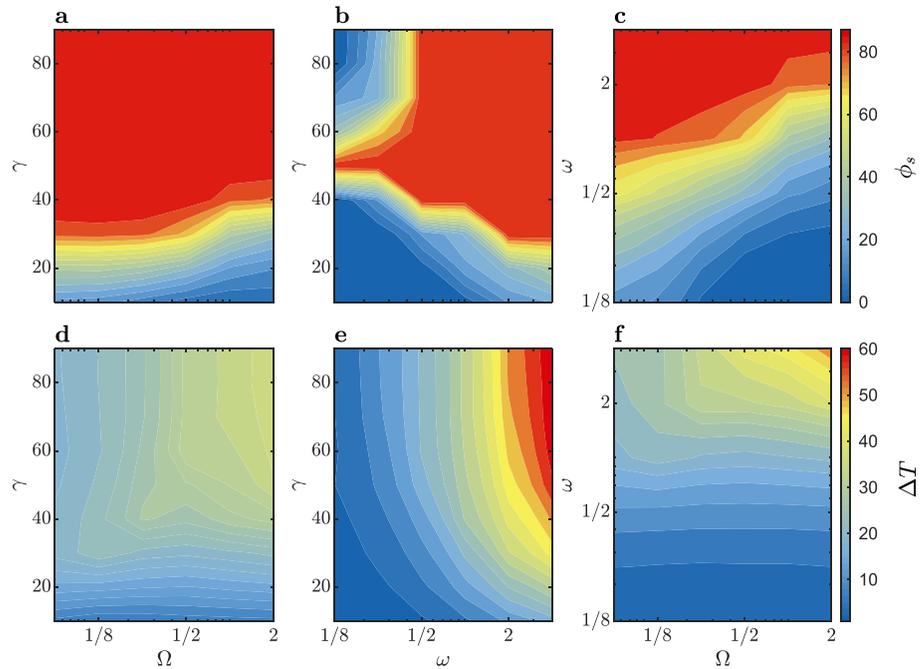

**Figure 3.** Seasonality indices, the latitudinal shift of maximum temperature, $\phi_s$ (degree of latitude, top row), and seasonal temperature amplitude, $\Delta T$ (K, bottom row), as a function of obliquity ($\gamma$), rotation rate ($\Omega$), and orbital period ($\omega$) for the general circulation model. In panels a and d, $\omega = 1$. In panels b and e, $\Omega = 1$. In panels c and f, $\gamma = 30°$.

Figure 3 confirms the description made in the previous paragraph in a more quantitative manner. The most dominant parameters that determine the seasonality are the obliquity and orbital period. Both $\phi_s$ and $\Delta T$ increase with increasing orbital period and obliquity for most cases. The only exception occurs for short orbital periods, where the annual mean climate dominates. In this case, $\phi_s$ changes non-monotonically, increasing toward $\gamma \approx 54°$ and then decreasing (Figure 2b). This is a result of this specific obliquity setting the transition from a normal to a reverse annual mean climate, that is, shifting from maximum to minimum temperature at the equator (e.g., Guendelman & Kaspi, 2019; Ohno & Zhang, 2019).

The rotation rate has a weaker and, as mentioned, a complex effect. Although having a weaker effect, it can become an influential factor when the obliquity values are intermediate, similar to the obliquity in a large number of the solar system planets, where $\phi_s$ can increase significantly as the rotation rate decreases (Figure 3c). Additionally, the contrasting effect of the rotation rate, where, as the rotation rate decreases $\Delta T$ decreases and $\phi_s$ increases, is an indication that the driving mechanism for this unique response relates to the meridional AHT. As the rotation rate decreases, the mean meridional circulation widens and strengthens, transporting more efficiently heat and lowering $\Delta T$ (e.g., Kaspi & Showman, 2015). At the same time, the latitude of the ascending branch is not confined to the latitude of maximum heating and can be poleward of it (Faulk et al., 2017; Guendelman & Kaspi, 2018; Hill et al., 2019, 2021). As a result, heat is transported poleward of the maximum heating, thereby shifting poleward the maximum temperature and increasing $\phi_s$.

### 3.2. Energy Balance Model

A diffusive EBM is an idealized model that considers energy balance with a highly idealized representation of meridional AHT through diffusion (e.g., Budyko, 1969; North, 1975; North & Coakley, 1979; North & Kim, 2017; Rose et al., 2017; Sellers, 1969; Williams & Kasting, 1997). We use the non-dimensional EBM presented in Section 2, which depends on three parameters, the obliquity ($\gamma$), the ratio between the radiative timescale and the orbital period ($\eta$), and the non-dimensional diffusion coefficient ($\mu$). As noted, the diffusivity in this model acts as a parameterization of the AHT, where an increase in $\mu$ means more efficient heat transport. A comparable physical parameter is the rotation rate: where as the rotation rate decreases the AHT becomes more efficient (e.g., Cox et al., 2021; Kaspi & Showman, 2015; Liu et al., 2017). Thus, decrease in the rotation rate can be considered to reflect an increase in $\mu$ (e.g., Cox et al., 2021; Liu et al., 2017; Williams & Kasting, 1997).

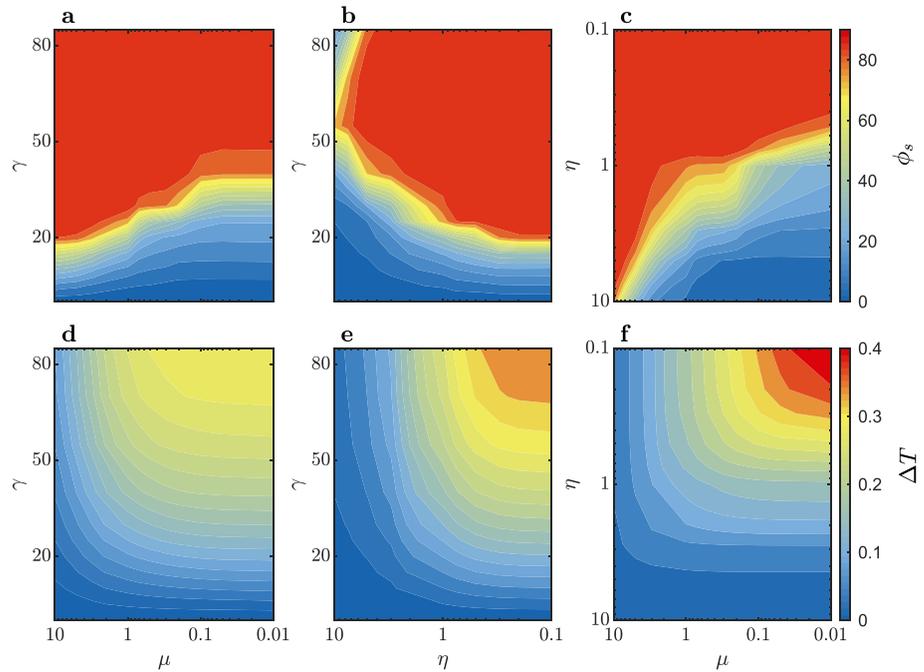

**Figure 4.** Seasonality indices, the latitudinal shift of maximum temperature, $\phi_s$ (degrees of latitude, top row), and the seasonal amplitude $\Delta T$ (non-dimensional temperature, bottom row), as a function of obliquity ($\gamma$), non-dimensional diffusion ($\mu$), and non-dimensional radiative timescale ($\eta$) for the energy balance model. In panels a and d, $\eta = 1$. In panels b and e, $\mu = 1$. In panels c and f, $\gamma = 30°$. Note that the axes directions of $\eta$ and $\mu$ are flipped (for easier comparison to Figure 3).

Evidently, the EBM results show great similarity to the GCM results, with the non-dimensional radiative timescale $\eta$ and diffusion $\mu$ having the inverse role of the orbital period ($\omega$) and rotation rate ($\Omega$), respectively (Figure 4, note the flipped directions of $\eta$ and $\mu$). Increasing the obliquity and decreasing $\eta$ (equivalent to increasing the orbital period relative to the radiative timescale) results in an increase in both $\phi_s$ and $\Delta T$ (Figure 4). In the EBM there is also an exception at large $\eta$, where $\phi_s$ changes non-monotonically with $\gamma$ (Figure 4b), since at high $\eta$ the annual mean climate dominates, this is a similar dependence for a short orbital period in the GCM (Figure 3b). Similar to the effect of decreasing the rotation rate in the GCM, increasing $\mu$ also results in a complex response. Increasing $\mu$, on the one hand, increases $\phi_s$ (Figures 4a and 4c) and, on the other hand, decreases $\Delta T$ (Figures 4d and 4f). These results strengthen the understanding that the effect of rotation rate on the climate seasonality stems from its effect on the AHT, that is, the increase in AHT efficiency with decreasing (increasing) rotation rate (diffusion).

The idealized framework of the EBM allows an easier inspection of the effect of AHT on seasonality. By decomposing the temperature field into a symmetric (the annual mean component, $T_{sym}$) and asymmetric (the seasonal component, $T_{asym}$) component, and comparing between the two, we show that AHT acts differently on each component, giving more insight to the effect of AHT on seasonality (Figure 5). At any given time, at the latitude of maximum temperature, by definition, we have $\partial_\phi T = 0$ or, alternatively, using the decomposition, $\partial_\phi T_{sym} = -\partial_\phi T_{asym}$. The left panel in Figure 5 shows the symmetric (solid lines) and asymmetric (dashed lines) parts of the meridional temperature gradient for three cases with a different value of $\mu$ but otherwise the same. This figure clearly depicts that increasing the diffusion acts more efficiently on the symmetric part (solid lines in the left panel and black line in the right panel) than on the asymmetric part (dashed lines in the left panel and blue line in the right panel). This differential effect causes the latitude where $\partial_\phi T_{sym} = -\partial_\phi T_{asym}$, which is also the latitude of maximum temperature, to be more poleward with increasing diffusion (dotted lines in the left panel of Figure 5).

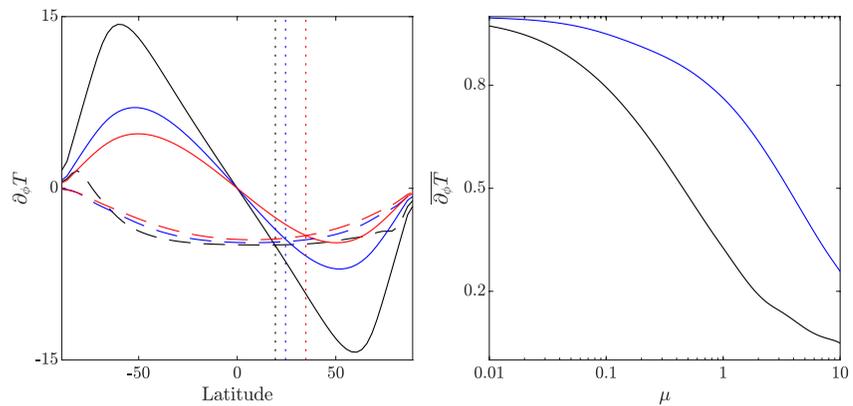

**Figure 5.** Left panel: $\partial_\phi T_{\text{sym}}$ (solid line) and $-\partial_\phi T_{\text{asym}}$ (dashed line) for $\gamma = 20°$, $\eta = 1$, and $\mu = 0.05$ (black), 0.5 (blue) and 1.0 (red). Dotted lines show the latitude of maximum temperature for each case, which is also the intersection of the two curves. Right panel: Normalized latitudinal mean of $\partial_\phi T_{\text{sym}}$ (black) and $\partial_\phi T_{\text{asym}}$ (blue) as a function of $\mu$ for $\gamma = 20°$, $\eta = 1$. The mean gradient is calculated by $\overline{\partial_\phi T} = \int_{-\pi}^{\pi} |\partial_\phi T|/2\pi$ and then normalized by the maximum value, that is, the value of $\overline{\partial_\phi T}$ for $\mu = 0$.

We can get an analytical relation for the dependence of the maximum temperature latitude on the diffusion, using a linear time-independent EBM (North & Kim, 2017)

$$D\nabla^2 T - BT = A - Q, \tag{14}$$

where the OLR is parameterized by $A + BT$, and $\nabla = \partial_x((1 - x^2)\partial_x)$, with a perpetual solstice-like case where the forcing now is given by:

$$Q = \frac{S_0}{4}\left[1 + \Delta_h \left(P_2(x) + 3\sin\phi_0 P_1(x)\right)\right], \tag{15}$$

where $P_1(x) = x$, $P_2(x) = (1 - 3x^2)/2$ are the first and second Legendre polynomials, with $x = \sin\phi$. $\Delta_h$ denotes the insolation meridional difference and $\phi_0$ the latitude of maximum forcing. This is a simplified, time-independent form for the approximation forcing of the seasonal cycle used in Rose et al. (2017). This can be considered an extreme case with an infinitely long orbital period. An analytical solution for this case is given by

$$T = \frac{S_0 \Delta_h}{4(6D + B)} P_2(x) + \frac{3S_0 \Delta_h \sin\phi_0}{4(2D + B)} P_1(x) + \frac{S_0 - 4A}{4B}. \tag{16}$$

This solution is consistent with the findings of Rose et al. (2017) in that it has similar coefficients for the Legendre polynomials. Here, the AHT or diffusion effects becomes clear, as the symmetric part is given by the $P_2(x)$ and the constant terms, and the asymmetric part is given by the $P_1(x)$ term. More specifically, it is easy to derive an expression for the latitude of maximum temperature, $\phi_{\text{max}}$,

$$\frac{\sin\phi_{\text{max}}}{\sin\phi_0} = \frac{6D + B}{2D + B}. \tag{17}$$

Noting that $\frac{6D+B}{2D+B} \geq 1$ means that $\phi_{\text{max}} > \phi_0$ for $D > 0$, that is, the latitude of the maximum temperature will be poleward of the latitude of maximum forcing for $D > 0$. Additionally $\phi_{\text{max}}$ will increase as the diffusion increases. This result emphasizes the role of AHT in determining the seasonal cycle a planet experiences. On the one hand, the AHT acts to lower the seasonal amplitude, that is, $\Delta T$, and, on the other hand, due to the differential effect of the AHT on the symmetric and asymmetric components, a more efficient AHT results in a poleward shift of $\phi_s$. Additionally, this relation emphasizes that the diffusion dependence is independent on the functional form of the OLR, and holds also for both linear and nonlinear forms for the OLR.

## 4. Discussion and Conclusions

In this study we demonstrate that in addition to the dependence of seasonality on the obliquity, orbital period and radiative timescale, it also depends on parameters that strongly affect the AHT, such as the rotation rate. We further show that this result consistently appears in both an idealized GCM when varying the rotation rate and in a diffusive EBM when varying the diffusivity. Unlike the seasonal dependence on the obliquity, radiative timescale and orbital period, its dependence on the diffusivity and rotation rate is more complex. Increasing (decreasing) the diffusivity (rotation rate), decreases the meridional temperature gradient and the seasonal amplitude. However, it allows the latitude of maximum temperature to shift more poleward (increase of $\phi_s$). The increase in $\phi_s$ is explained by the differential effect of the AHT on the symmetric and asymmetric parts of the temperature.

Our results are in agreement with previous studies of the dependence of the mean meridional circulation on planetary parameters. A specific example is the Hadley circulation, which is mostly responsible for the AHT in slow rotating planets and planets with strong seasonality (Kaspi & Showman, 2015; Lobo & Bordoni, 2020; Schneider & Bordoni, 2008). At fast rotation rates, the flow tends to arrange in zonal bands, and the Hadley circulation is constrained to lower latitudes. In this case, the AHT at higher latitudes is dominated mainly by the eddies and less by the mean flow. Even when neglecting the eddy influence, that is, considering an axisymmetric flow with off equatorial heating, the Hadley cell remains at low latitudes, and the ascending edge of the cell is equatorward of the maximum heating (e.g., Faulk et al., 2017; Guendelman & Kaspi, 2018; Singh, 2019). For slower rotation rates, an opposite situation occurs, and the ascending branch of the circulation goes poleward of the latitude of maximum heating (e.g., Guendelman & Kaspi, 2018). In this case, there is a flow in the boundary layer around the latitude of maximum heating that advects the heat poleward, resulting in the latitude of maximum temperature shifting poleward of the latitude of maximum heating. Additionally, the circulation's heat transport efficiency increases with decreasing rotation rate (e.g., Cox et al., 2021; Edson et al., 2011; Kaspi & Showman, 2015; Liu et al., 2017; Merlis & Schneider, 2010; Noda et al., 2017; Pierrehumbert & Hammond, 2019), thus reducing the seasonal amplitude ($\Delta T$).

Each of the explored parameters affects a different process. The obliquity dictates the incoming solar radiation seasonal cycle. The orbital period dictates the time period that the atmosphere has to adjust to the changes in the insolation. The rotation rate is a key parameter for the dynamics controlling the AHT, which we show to have a complex effect on the seasonal cycle. Although other parameters are important for the climate response and its seasonality, these three parameters represent the three most important processes that affect the seasonality. An important advantage in analyzing these parameters is that they can be varied directly, without including other effects, and thus making it easier to separate and diagnose their effect on the seasonality. This is unlike, for example, the atmospheric radiative timescale that depends on various parameters, such as the atmospheric mass and equilibrium temperature (Equation 1), and changes in the parameters that control the radiative timescale, affect the climate in additional ways, making the analysis more complex (Chemke & Kaspi, 2017; Colose et al., 2019; Goldblatt et al., 2009; Guendelman & Kaspi, 2019; Kaspi & Showman, 2015). Conversely, varying the orbital period, with all other parameters remaining constant, is comparable to changing the ratio between the radiative timescale and orbital period, similar to varying the non-dimensional parameter $\eta$ in the EBM.

This study sheds light on the observed seasonality of the solar system planets. One example is the seasonality on Earth, Mars, and Titan. Although, all three planets have an obliquity of ∼25°, they experience a significantly different seasonal cycle (Figure 1). Earth has the weakest seasonality, while Mars has the strongest. The weak seasonality on Earth is attributed mostly to the ocean coverage, that is, its high thermal inertia. The effect of the thermal inertia and the radiative timescale on the seasonality are a leading order effect. A planet with high thermal inertia, and thus a radiative timescale that is longer then the planet's orbital period, will, in general experience a weak seasonality. However, this study shows that the rotation rate of a planet also plays a significant role in determining the seasonality that a planet experiences. For example, an Earth-like planet with a slower rotation rate, might experience a significantly different seasonal cycle, where on one hand the warmest latitude will shift poleward, and on the other hand, it will experience a weaker seasonal amplitude. This also points to the role of the rotation rate in the seasonal cycle of Titan, allowing the maximum temperature to reach higher latitudes despite its very long radiative timescale, while keeping the seasonal amplitude weak (Figure 1). Another example is the seasonality, or lack thereof, on the fast-rotating ice giants. Even though their radiative timescale is comparable to their orbital period (Conrath et al., 1990), their fast rotation rate and large radius suggest that the AHT there is not very efficient, that is, low $\mu$, strengthening the assumption that the annual mean climate is the dominant one there.

The results of this study emphasize the importance of studying the climate's dependence on planetary parameters. In particular, when studying the atmospheres of planets outside the solar system, the consideration of temporal variations due to a seasonal cycle needs to also account for the effect of the dynamics and dynamical parameters such as the rotation rate.

## Conflict of Interest

The authors declare no conflicts of interest relevant to this study.

## Data Availability Statement

The GCM source code is available in https://www.gfdl.noaa.gov/idealized-moist-spectral-atmospheric-model-quickstart/, run file and output examples are available in https://doi.org/10.5281/zenodo.1442928. The EBM output is available in https://doi.org/10.5281/zenodo.6801615.

**Acknowledgments**

The authors thank the three reviewers for their helpful comments. The authors acknowledge the support from the Minerva Foundation and the Helen Kimmel Center of Planetary Science at the Weizmann Institute of Science.